\documentclass[11pt]{article}
\usepackage{amsfonts}
\usepackage{latexsym,amsmath}
\usepackage{amssymb,array}
\usepackage[symbol]{footmisc}
\usepackage{fancyhdr}
\usepackage[a4paper, portrait, margin=1.1811in]{geometry}
\usepackage{amsthm}
\usepackage{relsize}
\usepackage{graphicx}
\usepackage{lscape}
\usepackage{booktabs}
\usepackage{longtable}
\usepackage{xcolor} 
\usepackage{multirow}
\usepackage{threeparttable}
\usepackage[longnamesfirst,sort]{natbib}
\bibpunct[, ]{(}{)}{;}{a}{,}{,}

\usepackage[colorlinks, citecolor=cyan]{hyperref}

\begin{document}
	\title{\textbf{Modeling of Vertical Distribution of Suspended Sediment Concentration in Open Channel Turbulent Flows Using Fractional Differential Entropy}}
	\author{Poulami Paul\and Chanchal Kundu\footnote{\textit{Corresponding
				author e-mail}: \color{cyan}ckundu@rgipt.ac.in;
			\color{cyan}chanchal$_{-}$kundu@yahoo.com.}
		\and Department of Mathematical Sciences\\
		Rajiv Gandhi Institute of Petroleum Technology\\
		Jais 229 304, U.P., India}
	\date{July, 2025}
	\maketitle
	\begin{abstract}	
		Suspended sediment concentration and sediment transport heavily correlates to fluid behavior, thus proving it to be a lucrative field for exploration. Most of the existing deterministic and probabilistic methods proved to be complex with high computation cost. In this paper, we proposed a simpler yet accurate and cost effective concentration model using fractional entropy due to Ubriaco for continuous domain, termed as fractional differential entropy (FDE). We estimated the type I distribution of suspended sediment concentration along the vertical direction in open channels considering the dimensionless normalized concentration as a random variable and constructing an optimization problem using the FDE. The surface concentration is assumed to be zero throughout the study. We further validate our FDE based concentration distribution model through  regression and error analysis using some selected experimental and field data. The results are compared with the existing concentration models, which show the superiority of the proposed model with respect to the aspects considered under this study.
	\end{abstract}	
	\section{Introduction}
	Researchers and river engineers have conducted a wide range of studies on sediment-laden flows. These investigations focus on how fluid-sediment interactions are influenced by turbulence in open hydraulic channels. The studies consider sediments of varying sizes (small, medium, and coarse), which are being transported through rivers, irrigation canals, and estuaries. Sediment transport through these channels can be categorized into two classes- bed load and suspended load \citep{de}. The bed load, comprises of relatively coarser grained sediments, whereas the suspended load consists of fine grained particles. However, there is a rough margin separating the suspended load from the bed load. The height of this border line from the channel bed is termed as reference level \citep{de}. Among the studies related to sediment transport, a large number of works are primarily focused at predicting the vertical distributions of suspended sediment concentration in uniform sediment-laden turbulent flows through hydraulic channels. One of the goals of these studies is to monitor the quality of the fluid in natural hydraulic channels and manage all associated risks or disasters such as floods or degradation of marine health due to toxic sediment particles. \\
	\hspace*{0.2in} A multitude of deterministic methods were proposed to study the vertical profile of suspended sediment particles \citep[refer to][]{cha,d,k,kg,mgk}. 
	Among these classical approaches, we have a well-known method derived analytically by \cite{r} from the Prandtl-von Karman logarithmic velocity model and linear shear stress distribution by considering that the upward diffusion and the downward movement of sediment in suspension maintains an equilibrium. The Rouse equation is defined as:
	\begin{equation}
		\frac{c}{c_r} = \Bigg[\frac{(y_* - y)}{y}\frac{r}{(y_* - r)}\Bigg]^{R_0}, \label{1.1}
	\end{equation} where $ y_* $ = flow surface height from the bed, $ c_r = c|_{y=r = 0.05y_*}$ is the reference concentration at a height $ y = r $ from the channel bed and $ R_0 $ is the Rouse number. \\
	\hspace*{0.2in} On the other hand, these conventional methods for representing the vertical distribution of suspended sediment particles in turbulent sediment-laden flows have serious limitations. For instance, the Rouse equation fails to describe the sediment concentration distribution at or near the channel bed or at the water surface. Moreover, in case of shortage of data, limited or incomplete information and small size of the sample, the classical methods for estimating the suspension distribution become quite difficult and unreliable. These shortcomings prompted researchers to develop more general and probabilistic approaches to better capture sediment transport dynamics and assess key factors affecting the vertical concentration distribution of sediments, such as particle size, settling velocity, turbulent shear stress, fluid velocity, eddy viscosity and diffusion coefficients. 
	\cite{chiu,ch} applied Shannon entropy \citep{s} and the maximum entropy principle \citep{j,je} to develop probabilistic models that yield the least biased distribution of suspended sediment concentration (SSC). The use of entropy theory caught considerable attention from hydraulic engineers and researchers, owing to its advantages over traditional methods. Notably, one such advantage is that it enables the quantification of uncertainty in concentration distribution and offers efficient, cost-effective approaches for estimating key parameters in terms of statistical moments such as the mean concentration. The existing SSC models have successfully employed the Shannon, Renyi, Tsallis and fractional Wang entropies for uniform sediment-laden turbulent flows along vertical and horizontal cross sections of open channels \citep[see][for a detailed review]{aks,ak,chiu,chi,ch,cs,kms,km,l,sl}. The findings of these studies claim that the entropy-based SSC models can be considered as generalized probabilistic models for obtaining an efficient and reliable way of estimation of parameters and goodness of fit of a distribution to a given set of observations. Thus, it results in a unified approach to the estimation of frequency distributions, even with limited data compared to the traditional deterministic methods \citep{c}.\\ 
	\indent Aligned with these studies, the current work focuses specifically on developing more general suspended sediment concentration models, with the reference level defined as the lowest elevation at which concentration data is available. The fractional differential entropy (FDE) is particularly chosen to model the vertical distribution of suspended sediment concentration in open channels since it offers a fractional generalization of Shannon entropy. The FDE is an extension of the fractional entropy proposed by \cite{u} to continuous domain as defined by \cite{f}. Due to the presence of a fractional parameter in the entropy formula, this information measure features a non-additive property. This might imply non-extensivity, which can open pathways for its applicability to systems with long range correlations. Thus, it poses an open question regarding the types of applications for its successful implementation, where the Shannon entropy faces  limitations. This inspired us to check the possibility of its application in representing the non-linear interactions between the sediment particles and fluid in a better way.\\
	\indent Motivated by this, the objectives of the current study are listed as: (i) to derive a fractional entropy-based suspended sediment concentration distribution representing the sediment profile of turbulent flows along the vertical section from bed to the surface of an open channel, (ii) to validate our model with field and experimental data, (iii) to check for improvements in the estimated results than the previous models through error and regression analysis. Throughout the study, we have considered type I mode of sediment distribution where the suspension concentration is known to increase uniformly throughout the flow domain from the flow surface to a reference level near the bed, assuming zero surface concentration. 
	\section{Preliminary concepts}
	After \cite{chiu} who pioneered the use of Shannon entropy in hydrology, several other entropy-based approaches were introduced with the purpose of describing the distribution of hydraulic variables.
	To describe the sedimentation concentration distribution along the vertical column of open channels, \cite{chi} formulated an expression for the sediment concentration given by
	\begin{equation}
		\frac{c}{c_0} = \Bigg[\frac{1-\frac{y}{D}}{1+(e^M-1)\frac{y}{D}}\Bigg]^{\lambda'};~ \lambda' = \frac{\omega_s u_{max}(1-e^{-M})}{\beta u_* 2M},	\label{1.2}
	\end{equation}
	where $ \omega_s = $ settling velocity of sediment particles; $ u_* = $ shear velocity; $ M = $ entropy parameter that can be determined from the relation between the mean and maximum velocity along a vertical cross section as defined in \cite{chi}. But this model faced some severe setbacks due to a large number of parameters involved in the hypothesis of cumulative distribution function.\\
	\hspace*{0.2in} \cite{ch} came up with an alternative expression for the suspended sediment concentration (SSC) distribution using Shannon entropy with the aid of deterministic and probabilistic models. This model took into consideration the time-averaged sediment concentration as the r.v. expressed as:
	\begin{equation}
		\frac{c}{c_0} = \frac{1}{N} \log \Big[e^N - (e^N - 1)\frac{y}{D}\Big], \label{1.3}
	\end{equation}
	where $D =$ vertical depth of flow from channel bed to water surface; $ c_0 = $ sediment concentration at the bed at $ y=0; N = $ entropy parameter that can be obtained from the ratio of mean concentration of sediment $ c_m $ to maximum concentration $ c_0 $:
	\begin{equation}
		\frac{c_m}{c_0} = \frac{e^N}{e^N - 1} - \frac{1}{N}. \nonumber
	\end{equation}
	\hspace*{0.2in} Further, in order to improve the previous estimated results, \cite{cs} attempted to define the sediment concentration distribution in terms of Tsallis entropy by assuming the cumulative distribution function (cdf) to be a non-linear function of vertical distance from the channel bed. This is given by:
	\begin{equation}
		\frac{c}{c_0} = 1 - \frac{1}{N}\{1 - [(1 + 0.5 \log N)F(c) - 0.5 \log N]^{2/3}\}, \label{1.4}	
	\end{equation} 
	where $ F(c) $ is the hypothesized cdf.  $ N $ is a dimensionless entropy parameter as expressed in \cite{cs}.\\
	\hspace*{0.2in} Further, \cite{kms} gave a formulation for the suspended sediment concentration distribution in terms of R$\acute{e}$nyi entropy. This model assumed the cdf to be combination of a non-linear function of height and exponential function and considered non-zero concentration at the surface unlike the aforementioned models, so as to represent more realistic situations. It is expressed as:
	\begin{equation}
		\frac{c}{c_a} = \frac{\mu}{1-\mu} + \Bigg\{\Bigg(\frac{\mu}{\mu-1}+\hat{c}_D\Bigg)^{m/(m-1)}
		+\Bigg[\Bigg(\frac{2\mu - 1}{\mu-1}\Bigg)^{m/(m-1)} - \Bigg(\frac{\mu}{\mu-1}+\hat{c}_D\Bigg)^{m/(m-1)}\Bigg] [F(\hat{c})]^{(m-1)/m}   \label{1.5}	
	\end{equation}
	where $ m = $ entropy order whose value is taken to be $ 3/2$ and rest of the parameters are as defined in \cite{kms}. \\
	\hspace*{0.2in} Recently, \cite{aks} introduced the use of incomplete information-based fractional entropy proposed by Wang to derive SSC distribution of type I and type II. They formulated a new cumulative distribution function using the Mittag-Leffler function which described the experimental and field data in a better way than the previous entropy based-methods. It was also established that the fractional entropy was able to better capture the distribution of variables of such chaotic systems where information about the phase space is not completely available due to complex interactions in a turbulent flow.
	\section{Estimation of sediment concentration}
	Let us consider the time-averaged sediment concentration $ \mathcal{C} $ as a random variable (r.v.). For the type I  suspended sediment concentration distribution, $ \mathcal{C} $ assumes maximum value $ c_r $ at a reference level (say $ y_r $) near the channel bed and decreases with increase in height from $ y = y_r $ and assumes minimum value $ c_h, $ ($ c_h $ not necessarily zero) at the water surface at $ y=h $. 
	Therefore, the estimation of sediment concentration distribution along the vertical section of an open channel from $ y_r $ to $ h $ using FDE will involve the steps: 
	(i) definition of FDE,
	(ii) description of the constraint equations,
	(iii) defining the Lagrangian function by maximizing the entropy with respect to the constraints,
	(iv) determination of the Lagrange multipliers,
	(v) estimation of the probability density function, and
	(v) framing of hypothesis of cumulative distribution function for the determination of the distribution of suspended sediment concentration.
	
	\subsection{Definition of the fractional differential entropy}
	\cite{u} generalized the Shannon entropy by replacing the integer order derivative with the left Riemann-Liouville fractional derivative $ {}_a^{RL}D_t^{\alpha}, a = -\infty $ and $ 0 \leq \alpha < 1 $ in the expression
	\begin{equation}
		H_S(p) = \lim\limits_{t\to -1}\frac{d}{dt}\sum\limits_{i} p(i)^{-t}. \label{2.6}	
	\end{equation}
	Hence, the fractional order entropy proposed by Ubriaco is defined as
	\begin{equation}
		H_U^\alpha (p) = \sum_{i=1}^{n}p(i) (-\log p(i))^\alpha;~ \alpha \in [0,1] \label{2.7},
	\end{equation}
	where $ p(i) $ is the probability of the $ i^{th} $ value of the r.v. and $ (-\log p(i))^\alpha $ represents the fractional order information content related to the variable. When $ \alpha = 1 $, $ H_U^\alpha (p) $ becomes the Shannon entropy. 
	The fractional Ubriaco entropy (\ref{2.7}) when extended to the continuous domain is called the fractional differential entropy (FDE). To the best of our knowledge, this fractional entropy has not been explored much in any physical domain due to the presence of the fractional order information content $ (-\log f(x))^\alpha $ of the continuous r.v $ X $ having probability density function (pdf) $ f $. This leaves it an open challenge to describe the physical processes in terms of FDE.  
	Henceforth, if we consider the normalized concentration $ \hat{\mathcal{C}} = \frac{1}{c_r}\mathcal{C} $ as the r.v. in continuous domain $ [0,1], $ then the FDE of $ \hat{\mathcal{C}} $ can be expressed as 
	\begin{equation}
		H^\alpha (f) = \int_{\hat{c}_h}^{1} f(\hat{c})(-\log f(\hat{c}))^\alpha~ d\hat{c}, \label{2.8}
	\end{equation}
	where $ \hat{c}_h $ is the dimensionless concentration at the flow surface $ y=h $. 
	\subsection{Description of the constraint equations}
	The constraint equations that need to be satisfied by the concentration distribution can be expressed as: 
	\begin{eqnarray}
		\int_{\hat{c}_h}^{1} f(\hat{c})~d\hat{c} = 1 & \label{2.9}\\
		\int_{\hat{c}_h}^{1} \hat{c} f(\hat{c})~d\hat{c} = \hat{c}_m ~,&  \label{2.10}
	\end{eqnarray}
	where $ \hat{c}_m = \frac{c_m}{c_r} $ is the mean sediment concentration of the vertical channel cross-section. \\
	\hspace*{0.2in} The constraint equation (\ref{2.9}) represents the total probability law or the normalization condition that is satisfied by every distribution. The constraint equation (\ref{2.10}) defines the law of conservation of mass of sediment. However, there are infinitely many distributions satisfying the constraints (\ref{2.9}) and (\ref{2.10}). So, to find the desired distribution for the vertical distribution of suspended sediment concentration (SSC), we need to maximize the entropy (\ref{2.8}) using the concept of POME \citep{j}. More constraint equations can be included but for simplicity of the model without loosing its prediction accuracy, only two constraints are used \citep{aks}.
	
	\subsection{Defining the Lagrangian function to maximize the entropy}
	Using the Lagrange multiplier method, the entropy (\ref{2.8}) is maximized subject to the constraints (\ref{2.9}) and (\ref{2.10}) to obtain the most suitable pdf $f(\hat{c}$). The Lagrangian function $ L $ can be written as:
	\begin{equation}
		L = \int_{\hat{c}_h}^{1} f(\hat{c})(-\log f(\hat{c}))^\alpha~ d\hat{c} + \lambda_0 \Big[\int_{\hat{c}_h}^{1} f(\hat{c})~d\hat{c} - 1\Big] + \lambda_1 \Big[\int_{\hat{c}_h}^{0} \hat{c} f(\hat{c})~d\hat{c} - \hat{c}_m\Big] \label{2.11}
	\end{equation} 
	where $ \lambda_0 $ and $ \lambda_1 $ are the Lagrange multipliers. 
	Solving the Euler-Lagrange equation treating $ f(\hat{c}) $ and $ \hat{c} $ as the dependent and independent variables, respectively, given by
	\begin{equation*}
		\frac{\partial L}{\partial f(\hat{c})} - \frac{d}{d\hat{c}}\Big(\frac{\partial L}{\partial f(\hat{c})'}\Big) = 0,
	\end{equation*} and ignoring the integral sign, we have
	\begin{equation}
		\frac{\partial f(\hat{c})[-\log f(\hat{c})]^\alpha}{\partial f(\hat{c})} ~+~ \lambda_0~ \frac{\partial f(\hat{c})}{\partial f(\hat{c})} ~+~ \lambda_1~\frac{\partial \hat{c} f(\hat{c})}{\partial f(\hat{c})} = 0.\label{2.12}
	\end{equation}
	Simplifying (\ref{2.12}) gives us
	\begin{equation}
		\alpha(-\log f(\hat{c}))^{\alpha - 1} - (-\log f(\hat{c}))^\alpha + \lambda_0 + \lambda_1 \hat{c} = 0. \label{2.13}
	\end{equation}
	Solving (\ref{2.13}) for $ \alpha = \frac{1}{2}, $ the pdf can be derived as:
	\begin{equation}
		f(\hat{c}) = \exp\Bigg[-\frac{(1+(\lambda_0+\lambda_1 \hat{c})^2) \pm (\lambda_0+\lambda_1 \hat{c}) \sqrt{(\lambda_0+\lambda_1 \hat{c})^2 + 2}}{2}\Bigg]. \label{2.14}
	\end{equation}
	Eq. (\ref{2.14}) is the estimated entropy based pdf of suspended sediment concentration, which can be used to identify the distribution of sediment concentration in open channels. 
	Substituting the value of $ f(\hat{c}) $ from (\ref{2.14}) in (\ref{2.8}), we obtain the maximum entropy of $ f(\hat{c}) $ of $ \hat{\mathcal{C}} $.
	Therefore, we have the fractional order differential entropy which can be computed as follows:
	\begin{align}
		H^\alpha(f) = \int_{\hat{c}_h}^{1} \exp\Bigg[-\frac{(1+(\lambda_0+\lambda_1 \hat{c})^2)\pm (\lambda_0+\lambda_1 \hat{c}) \sqrt{(\lambda_0+\lambda_1 \hat{c})^2 + 2}}{2}\Bigg]\times \nonumber\\
		\Bigg(\frac{(1+(\lambda_0+\lambda_1 \hat{c})^2)\pm (\lambda_0+\lambda_1 \hat{c}) \sqrt{(\lambda_0+\lambda_1 \hat{c})^2 + 2}}{2}\Bigg)^\alpha ~d\hat{c}. \label{2.15i}
	\end{align}
	The entropy function (\ref{2.15i}) can be numerically solved for a given data set once the Lagrange multipliers are computed for the data.
	
	\subsection{Determination of cumulative distribution function}
	The vertical distribution of SSC can be obtained by finding a generalized cumulative distribution function (cdf) in terms of the concentration domain and relating it with that of the spatial domain. This relation can be established by considering the fact that the cdf is a decreasing function of the spatial variable $ y/h $ and increases with increase of the considered random variable of the flow domain $ \hat{c}. $
	Further, the cdf is supposed to assume the maximum value $ 1 $ at the reference level $ y= y_r $ and attains the minimum value $ 0 $ at the channel surface $ y = h. $ Therefore considering the geometry of the flow space laden with suspended sediments and the above assumptions, the cdf of sediment concentration $ F(\hat{c}) $ can be framed as
	\begin{equation}
		F(\hat{c}) = \Big(1 - \Big(\frac{y-y_r}{h-y_r}\Big)^a\Big) \exp\Big(- \Big(\frac{y-y_r}{h-y_r}\Big)^a\Big), \label{2.15}
	\end{equation}
	where $ a $ can be obtained from a given set of observations and is related to the characteristics of the sediment particles \citep{kms}. 
	The values of $ a \in (0,1) $ indicates the rate of decline of the cdf from the channel bed region to the water surface. Smaller $ a $ values indicate slow decrease in cdf while larger $ a $ values will lead to faster decrease in cdf \citep{cs}.  \\
	Now, the general form of cdf is derived by integrating the entropy based pdf (\ref{2.14}) expressed as: 
	\begin{equation}
		F(\hat{c}) = \int_{\hat{c}_h}^{\hat{c}} exp\Bigg[-\frac{(1+(\lambda_0+\lambda_1 \hat{c})^2)\pm (\lambda_0+\lambda_1 \hat{c}) \sqrt{(\lambda_0+\lambda_1 \hat{c})^2 + 2}}{2}\Bigg]~d\hat{c} \label{2.16}  
	\end{equation}
	Taking care of the decreasing nature of the cdf in terms of height from the reference level and increasing nature in terms of concentration $ \hat{c} $, we will take only the case where we have 
	\begin{equation}
		f(\hat{c}) = exp\Bigg[-\frac{(1+(\lambda_0+\lambda_1 \hat{c})^2)- (\lambda_0+\lambda_1 \hat{c}) \sqrt{(\lambda_0+\lambda_1 \hat{c})^2 + 2}}{2}\Bigg] \label{2.17}
	\end{equation}
	Using the value of $ f(\hat{c}) $ from eq. (\ref{2.17}) and equating (\ref{2.16}) with (\ref{2.15}), we obtain 
	\begin{equation}
		\int_{\hat{c}_h}^{\hat{c}} exp\Bigg[-\frac{(1+(\lambda_0+\lambda_1 \hat{c})^2) - (\lambda_0+\lambda_1 \hat{c}) \sqrt{(\lambda_0+\lambda_1 \hat{c})^2 + 2}}{2}\Bigg]~d\hat{c} = (1 - \hat{Y}^a)\cdot\exp(-\hat{Y}^a) \label{2.18}
	\end{equation}
	where $ \hat{Y} = \mathlarger{\frac{y-y_r}{h-y_r}} .  $
	For computational convenience, let us assume $ \hat{c_h} = 0 $ at the surface of the flow. Solving the integral on the L.H.S of eq. (\ref{2.18}) by using the substitution $ (\lambda_0 + \lambda_1 \hat{c})^2 = t $, we get
	\begin{equation}
		\frac{e^{-1/2}}{2\lambda_1} \int_{\lambda_0^2}^{(\lambda_0+\lambda_1\hat{c})^2} t^{-1/2} e^{-\frac{1}{2}(t-\sqrt{t^2 + 2t})}~dt \label{2.19}
	\end{equation}
	By using the identity $ e^{-x} = \sum_{i=0}^{\infty} \frac{(-1)^i}{i!} x^i, $
	(\ref{2.19}) gives
	\begin{equation}
		\frac{e^{-1/2}}{2\lambda_1} \int_{\lambda_0^2}^{(\lambda_0+\lambda_1\hat{c})^2} t^{-1/2} \sum_{i=0}^{\infty} \frac{(-1)^i}{i!} \Bigg(\frac{t- \sqrt{t^2 + 2t}}{2}\Bigg)^i~dt \label{2.20}
	\end{equation}
	On using binomial expansion $ (x-y)^n = \sum_{i=o}^{n}(-1)^i {n \choose i}  x^{n-i}y^i $ in (\ref{2.20}), we get
	\begin{equation}
		\frac{e^{-1/2}}{2\lambda_1}\sum_{i=0}^{\infty} \sum_{j=0}^{i} \frac{(-1)^{i+j}}{2^i i!} {i \choose j} \int_{\lambda_0^2}^{(\lambda_0+\lambda_1)^2} t^{-1/2} t^{i-j}(t^2 + 2t)^{j/2}~dt	\label{2.21}
	\end{equation}
	Again, using the the identity  $ (1+x)^j = \sum_{k=0}^{\infty}{j \choose k} x^k,~j\in \mathcal{R} $ and the assumption $ |t| < 1, $ (\ref{2.21}) can further be reduced to
	\begin{equation}
		\frac{e^{-1/2}}{\lambda_1}\sum_{i=0}^{\infty} \sum_{j=0}^{i} \sum_{k=0}^{\infty} \frac{(-1)^{i+j}}{i!} {i \choose j} {j/2 \choose k} 2^{j/2 - i - k} \Bigg[\frac{(\lambda_0 + \lambda_1 \hat{c})^{1+2(i+k)-j}- \lambda_0^{1+2(i+k)-j}}{1+2(i+k)-j}\Bigg] 	\label{2.22}	
	\end{equation}
	Ignoring the higher order terms of (\ref{2.22}), for the case when $ i=0,1; k= 0, $ we obtain
	\begin{equation}
		F(\hat{c}) = \frac{e^{-1/2}}{2\sqrt{2} \lambda_1} ((\lambda_0 + \lambda_1 \hat{c})^2 - \lambda_0^2) \label{2.23}
	\end{equation}
	Therefore, from eqs. (\ref{2.18}) and (\ref{2.23}), we obtain an explicit expression for $ \hat{c} $ given by
	\begin{equation}
		\hat{c} = \frac{1}{\lambda_1} \Bigg[-\lambda_0 \pm \sqrt{\lambda_0^2 + \lambda_1 \Big(1 - \Big(\frac{y-y_r}{h-y_r}\Big)^a\Big) \exp\Big(- \frac{y-y_r}{h-y_r}\Big)^a\Big)2\sqrt{2}e^{1/2}} \Bigg].  \label{2.24}
	\end{equation}
	Multiplying  eq. (\ref{2.24}) by the maximum concentration $ c_r $ will represent the distribution of suspended sediment concentration in turbulent flows along the vertical depth of open channels using fractional differential entropy, i.e., $ c = c_r\cdot\hat{c}. $
	
	\subsection{Determination of Lagrange multipliers}
	So, we are left with two unknowns $ \lambda_0 $ and $ \lambda_1 $ in eq. (\ref{2.24}) which needs to be determined. The values of the unknown Lagrange multipliers can be obtained by substituting $ f(\hat{c}) $ from eq. (\ref{2.17}) in the constraints (\ref{2.9}) and (\ref{2.10}). \\
	\hspace*{0.2in}By putting the value of $ f(\hat{c}) $ in the constraint (\ref{2.9}) and by using the substitution $ (\lambda_0 + \lambda_1 \hat{c})^2 = t $, we have
	\begin{eqnarray}
		L.H.S &=& \int_{\hat{c}_h}^{1} f(\hat{c})~d\hat{c} \nonumber\\
		&=& \frac{1}{2\lambda_1}\int_{(\lambda_0+\lambda_1\hat{c}_h)^2}^{(\lambda_0+\lambda_1)^2}
		\exp\Big[-\Big(\frac{1+t}{2}\Big)\Big]\cdot\exp\Big[\frac{\sqrt{2t+t^2}}{2}\Big]\frac{dt}{\sqrt{t}} \label{2.25} 
	\end{eqnarray}
	Now, using the power series expansion of $ e^{-x} $ in (\ref{2.25}), we get 
	\begin{eqnarray*}
		L.H.S &=& \frac{e^{-1/2}}{2\lambda_1}\int_{(\lambda_0+\lambda_1\hat{c}_h)^2}^{(\lambda_0+\lambda_1)^2} t^{-1/2}\sum_{i=0}^{\infty}\frac{(-1)^i}{i!}\Bigg(\frac{t-\sqrt{2t+t^2}}{2}\Bigg)^i dt.	
	\end{eqnarray*}
	Again, using binomial expansion formulae for expanding the series $ (x-y)^i,~i\in \mathcal{N} $ and $ (1+x)^i,~i\in \mathcal{R} $ for $ |x| < 1 $, we obtain
	\begin{eqnarray*}
		L.H.S &=& \frac{e^{-1/2}}{2\lambda_1}\sum_{i=0}^{\infty}\frac{(-1)^{i+j}}{2^i i!}\sum_{j=0}^{i}{i \choose j}\int_{(\lambda_0+\lambda_1\hat{c}_h)^2}^{(\lambda_0+\lambda_1)^2} t^{-1/2}t^{i-j}(2t)^{j/2}\Big(1+\frac{t}{2}\Big)^{j/2} dt; ~|t| < 1 \\ 
		&=& \frac{e^{-1/2}}{2\lambda_1}\sum_{i=0}^{\infty}\sum_{j=0}^{i}\sum_{k=0}^{\infty}\frac{(-1)^{i+j}}{2^i i!}{i \choose j}2^{\frac{j}{2}-k}{j/2 \choose k}\int_{(\lambda_0+\lambda_1\hat{c}_h)^2}^{(\lambda_0+\lambda_1)^2} t^{-\frac{1}{2}+i-\frac{j}{2}+k} dt \\
		&=& \frac{e^{-1/2}}{\lambda_1}\sum_{i=0}^{\infty}\sum_{j=0}^{i}\sum_{k=0}^{\infty}{i \choose j}{j/2 \choose k}\frac{(-1)^{i+j} 2^{\frac{j}{2}-k-i}}{i!} \Bigg[\frac{(\lambda_0+\lambda_1)^{1+2i-j+2k}-(\lambda_0+\lambda_1\hat{c}_h)^{1+2i-j+2k}}{1+2i-j+2k}\Bigg]. 
	\end{eqnarray*}
	Now, after assuming $ \hat{c}_h = 0 $, L.H.S = 1 yields:
	\begin{equation}
		\sum_{i=0}^{\infty}\sum_{j=0}^{i}\sum_{k=0}^{\infty}{i \choose j}{j/2 \choose k}\frac{(-1)^{i+j} 2^{\frac{j}{2}-k-i}}{i!} \Bigg[\frac{(\lambda_0+\lambda_1c_0)^{1+2i-j+2k}-\lambda_0^{1+2i-j+2k}}{1+2i-j+2k}\Bigg] = \lambda_1e^{1/2}. \label{2.26}
	\end{equation}
	For simplicity of the model, since we have taken $ |t|<1 \implies |\lambda_0+\lambda_1\hat{c}| < 1 $, we can assume $ |\lambda_0| < 1 ~\text{and}~ |\lambda_0+\lambda_1| < 1 $. Thus, we can approximate (\ref{2.26}) up to two terms by ignoring the higher power terms of $ (\lambda_0+\lambda_1) $ and $ \lambda_0. $ 
	Through this, we get a linear equation in $ \lambda_0 $ and $ \lambda_1 $.\\
	For the case: $(i=0,1; k=0) $
	\begin{equation}
		2\lambda_0 + \lambda_1 = 2\sqrt{2}e^{1/2}. \label{2.27}
	\end{equation}
	\hspace*{0.2in}Similarly, by using the value of $ f(\hat{c}) $ in the constraint (\ref{2.10}) and on using the substitution $ \lambda_0+\lambda_1\hat{c} = t $, we have
	\begin{eqnarray}
		L.H.S &=& \int_{\hat{c}_h}^{1} \hat{c}f(\hat{c})~d\hat{c} \nonumber\\
		&=& \frac{1}{\lambda_1}\int_{\lambda_0+\lambda_1\hat{c}_h}^{\lambda_0+\lambda_1} \Big(\frac{t-\lambda_0}{\lambda_1}\Big)
		\exp\Big[-\Big(\frac{1+t^2}{2}\Big)\Big]\cdot\exp\Big[\frac{\sqrt{2t^2+t^4}}{2}\Big]~dt \nonumber \\
		&=& \frac{e^{-1/2}}{\lambda_1^2}\Bigg[\int_{\lambda_0+\lambda_1\hat{c}_h}^{\lambda_0+\lambda_1}t\cdot\exp\Big[-\frac{t^2-\sqrt{2t^2+t^4}}{2}\Big]~dt - \lambda_0\int_{\lambda_0+\lambda_1\hat{c}_h}^{\lambda_0+\lambda_1}\exp\Big[-\frac{t^2-\sqrt{2t^2+t^4}}{2}\Big]~dt\Bigg]. \nonumber\\
		\text{Now},~ I_1 &=& \int_{\lambda_0+\lambda_1\hat{c}_h}^{\lambda_0+\lambda_1}t\cdot\exp\Big[-\frac{t^2-\sqrt{2t^2+t^4}}{2}\Big]~dt \nonumber\\
		&=& \sum_{i=0}^{\infty}\frac{(-1)^{i+j}}{2^i i!}\sum_{j=0}^{i} {i \choose j}\int_{\lambda_0+\lambda_1\hat{c}_h}^{\lambda_0}t^{1+2(i-j)}(2t^2+t^4)^{j/2}~dt. \nonumber 
	\end{eqnarray}
	To arrive at an explicit expression, let us assume $ |t|< 1 $.
	\begin{eqnarray}
		\implies I_1 &=& \sum_{i=0}^{\infty}\frac{(-1)^{i+j}}{2^i i!}\sum_{j=0}^{i} {i \choose j}\int_{\lambda_0+\lambda_1\hat{c}_h}^{\lambda_0}t^{1+2(i-j)+j}\cdot2^{j/2}\Big(1+\frac{t^2}{2}\Big)^{j/2}~dt ;~ \Big|\frac{t^2}{2}\Big| < 1 \nonumber \\
		&=& \sum_{i=0}^{\infty}\frac{(-1)^{i+j}}{2^i i!}\sum_{j=0}^{i} {i \choose j}\sum_{k=0}^{\infty} {j/2 \choose k} 2^{j/2 - k}\int_{\lambda_0+\lambda_1\hat{c}_h}^{\lambda_0}t^{1+2i-j+2k}~dt \nonumber \\
		&=& \sum_{i=0}^{\infty}\sum_{j=0}^{i}\sum_{k=0}^{\infty}{i \choose j} {j/2 \choose k}\frac{(-1)^{i+j} ~2^{j/2 - i - k}}{i!}\Bigg[\frac{(\lambda_0+\lambda_1)^{2(1+i+k)-j} - (\lambda_0+\lambda_1\hat{c}_h)^{2(1+i+k)-j}}{{2(1+i+k)-j}}\Bigg]. \nonumber\\
		\label{2.28}\\
		\text{Again,} ~ I_2 &=& \nonumber \int_{\lambda_0+\lambda_1\hat{c}_h}^{\lambda_0}\exp\Big[-\frac{t^2-\sqrt{2t^2+t^4}}{2}\Big]~dt \\
		&=& \sum_{i=0}^{\infty}\frac{(-1)^{i+j}}{2^i i!}\sum_{j=0}^{i} {i \choose j}\int_{\lambda_0+\lambda_1\hat{c}_h}^{\lambda_0}t^{2(i-j)+j}\cdot2^{j/2}\Big(1+\frac{t^2}{2}\Big)^{j/2}~dt \nonumber \\
		&=& \sum_{i=0}^{\infty}\sum_{j=0}^{i}\sum_{k=0}^{\infty}{i \choose j} {j/2 \choose k}\frac{(-1)^{i+j} ~2^{j/2 - i - k}}{i!}\Bigg[\frac{(\lambda_0+\lambda_1)^{1+2(i+k)-j} - (\lambda_0+\lambda_1\hat{c}_h)^{1+2(i+k)-j}}{{1+2(i+k)-j}}\Bigg]. \nonumber	\\\label{2.29}
	\end{eqnarray}
	Hence, combining (\ref{2.28}) and (\ref{2.29}) will give\\
	$ L.H.S = \frac{e^{-1/2}}{\lambda_1^2}[I_1 -\lambda_0I_2] = R.H.S = \hat{c}_m $ implies
	\begin{align}
		\sum_{i=0}^{\infty}\sum_{j=0}^{i}\sum_{k=0}^{\infty}{i \choose j} {j/2 \choose k}\frac{(-1)^{i+j} ~2^{j/2 - i - k}}{i!} \cdot \Bigg[\Bigg(\frac{(\lambda_0+\lambda_1)^{2(1+i+k)-j} - (\lambda_0+\lambda_1\hat{c}_h)^{2(1+i+k)-j}}{{2(1+i+k)-j}}\Bigg) \nonumber \\ 
		-\lambda_0\Bigg(\frac{(\lambda_0+\lambda_1)^{1+2(i+k)-j} - (\lambda_0+\lambda_1\hat{c}_h)^{1+2(i+k)-j}}{{1+2(i+k)-j}}\Bigg)\Bigg] = \lambda_1^2 e^{1/2}\hat{c}_m. \label{2.30}
	\end{align}
	Approximating up to 2 terms in (\ref{2.30}), we work out the cases for $(i=0,1; k=0)$: 
	\begin{equation}
		3\lambda_0 + 2\lambda_1 = 6\sqrt{2}e^{1/2}\hat{c}_m. \label{2.31}
	\end{equation}
	The Lagrange multipliers $\lambda_0$ and $\lambda_1$ can be obtained by solving the system of linear simultaneous equations expressed by (\ref{2.27}) and (\ref{2.31}).
	\section{Validation from experimental and field data}
\begin{figure}[h]
		\includegraphics[width=0.45\textwidth]{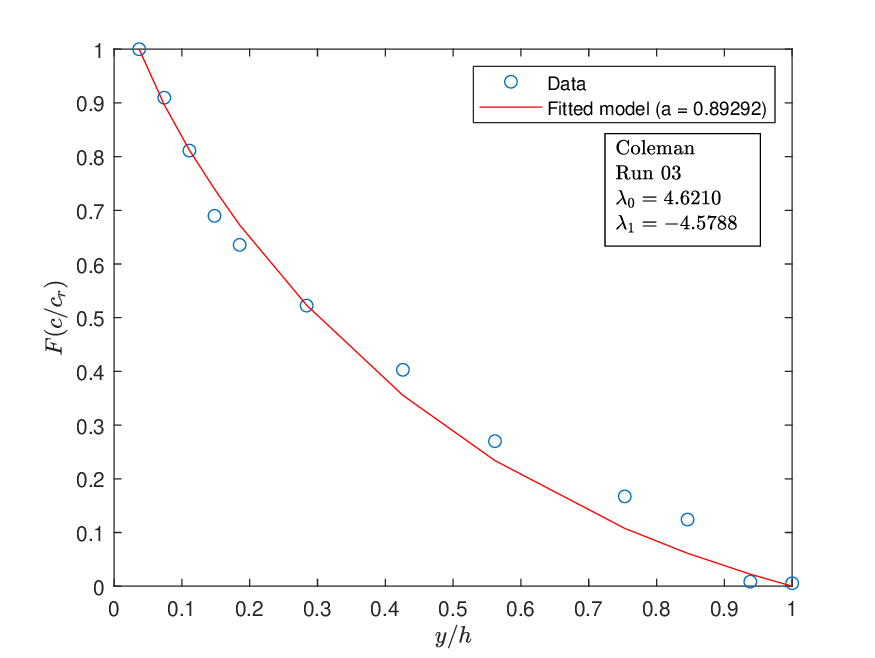} ~~~
		\includegraphics[width=0.45\textwidth]{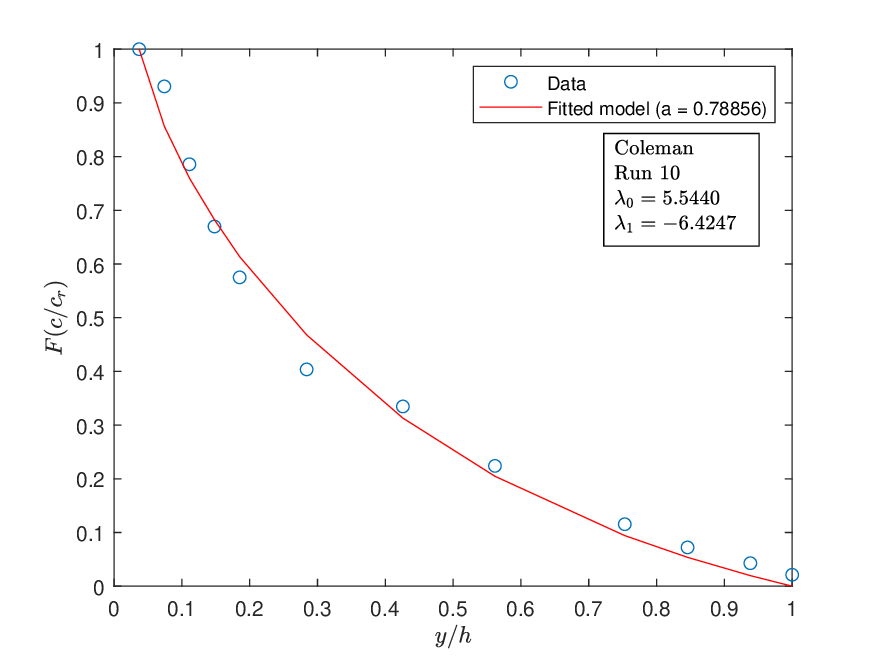}~~~\\
		\includegraphics[width=0.45\textwidth]{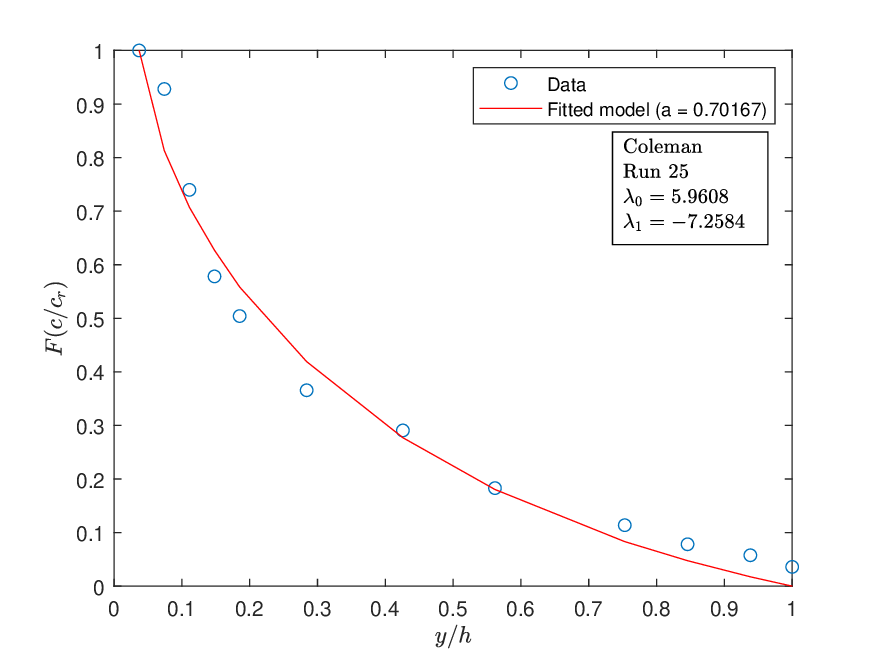}~~~
		\includegraphics[width=0.45\textwidth]{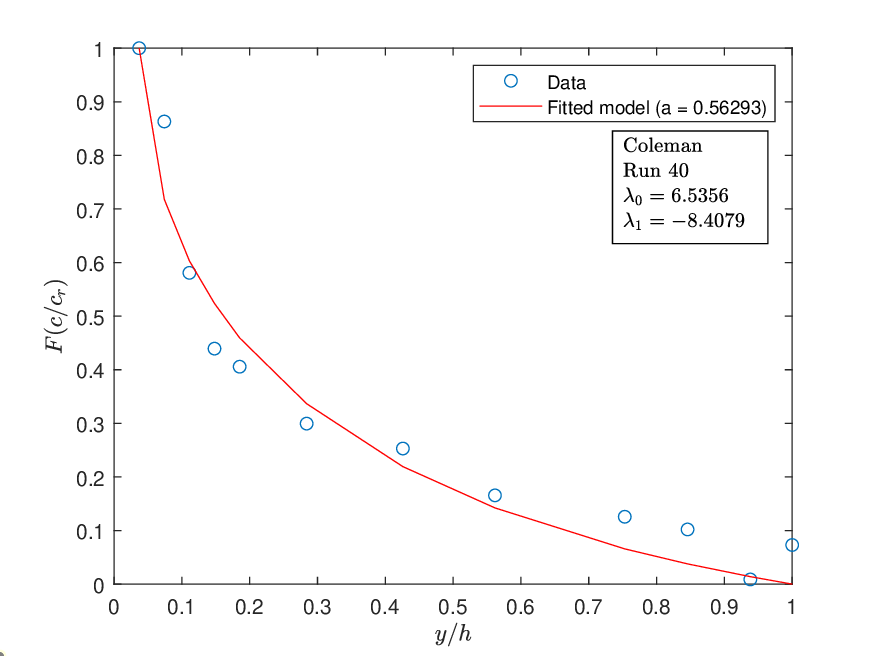}~~~ \\
		\includegraphics[width=0.45\textwidth]{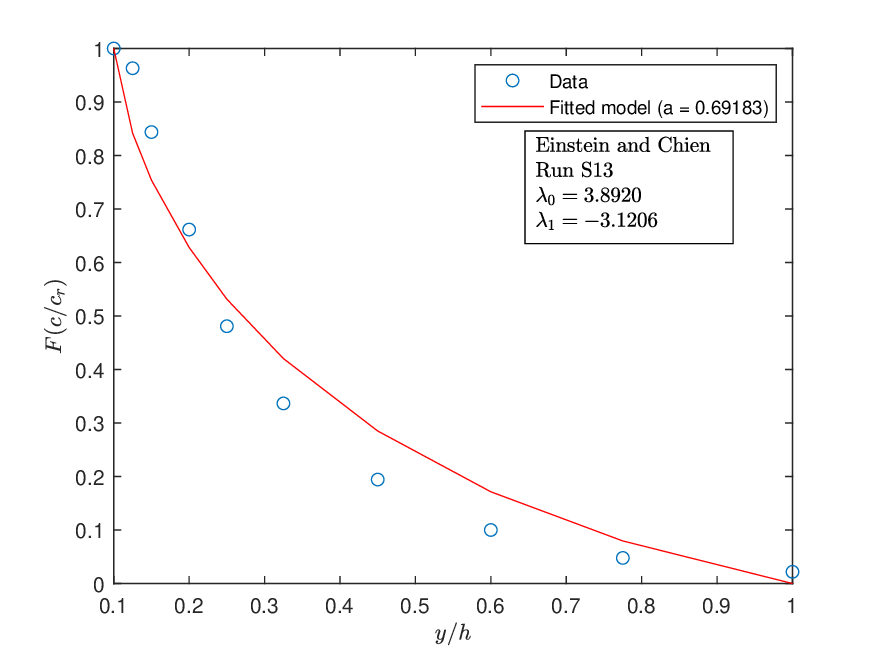}~~~ 
		\includegraphics[width=0.45\textwidth]{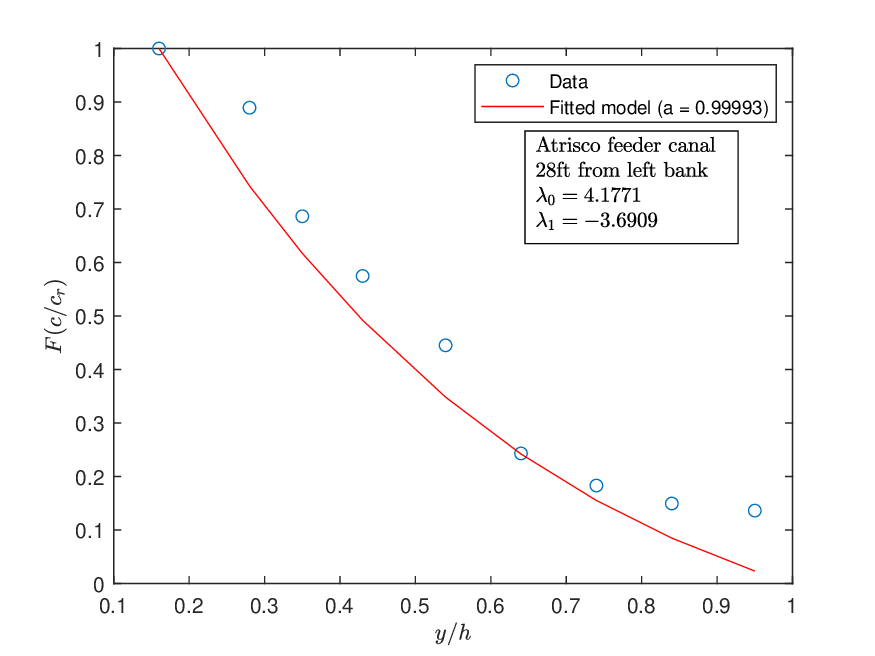}~~~ \\
		\includegraphics[width=0.45\textwidth]{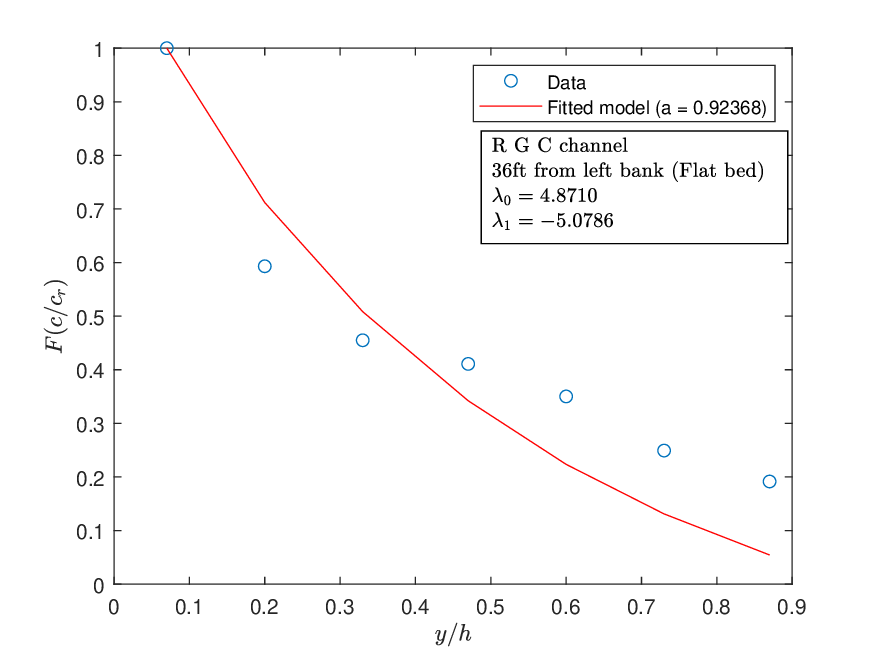}~~~ 
		\includegraphics[width=0.45\textwidth]{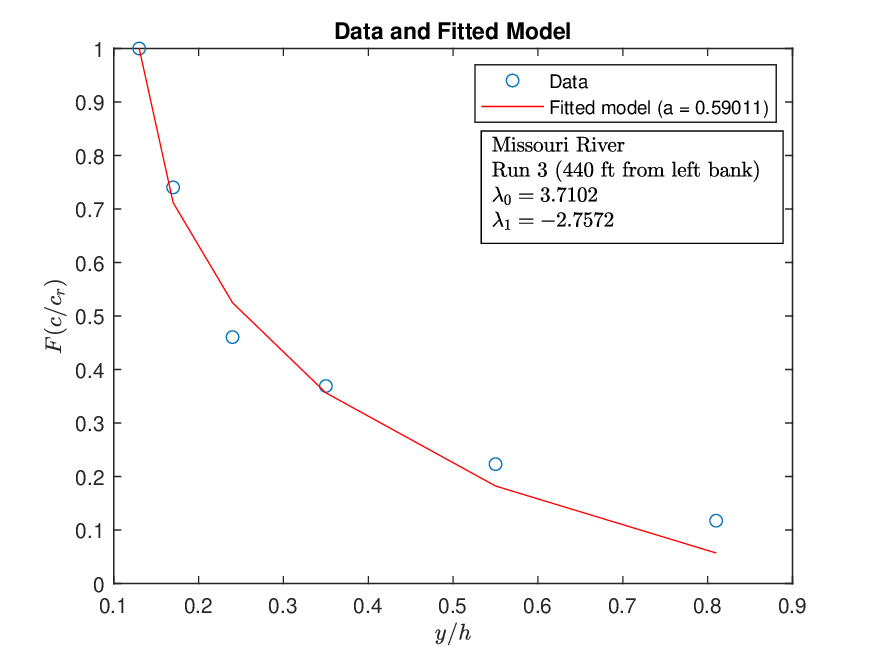}~~~
		\caption{Validation of the fitted hypothetical cdf (\ref{2.15}) with the estimated cdf (\ref{2.23}) for Coleman, Einstein \& Chien and McQuivey data}
		\label{cdf_fit}
	\end{figure}

\begin{figure}[]
		\includegraphics[width=0.45\textwidth]{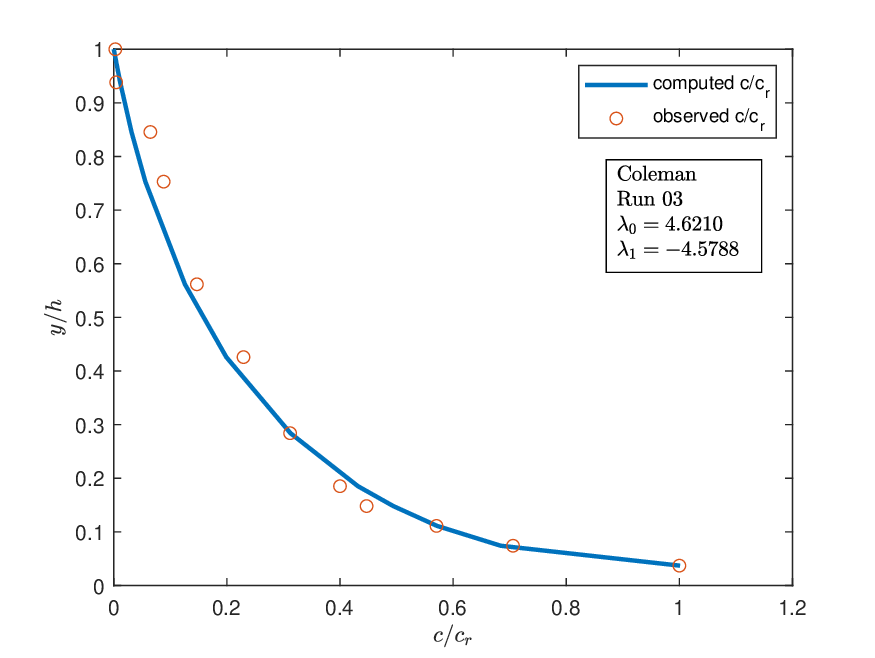} ~~~
		\includegraphics[width=0.45\textwidth]{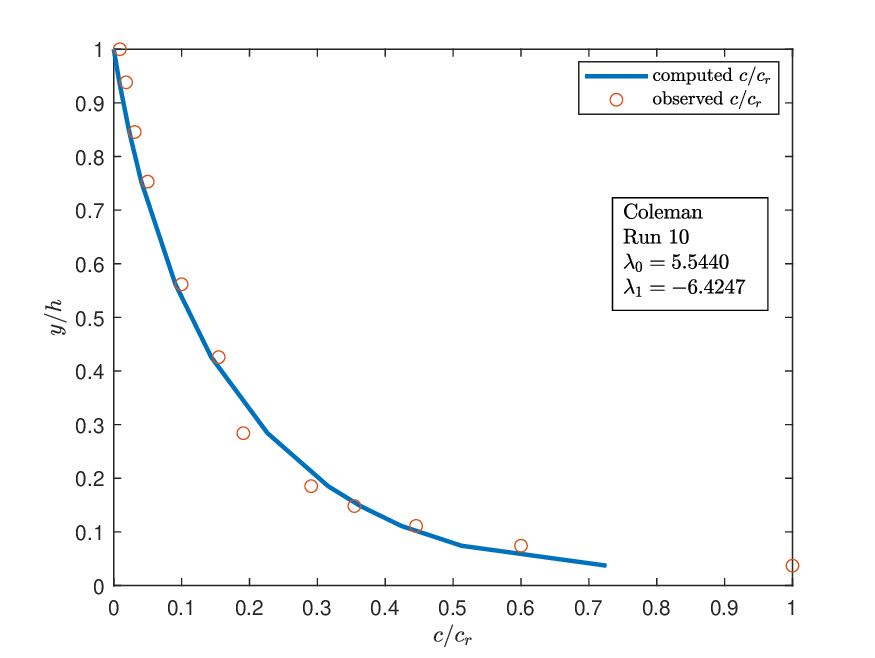}\\
		\includegraphics[width=0.45\textwidth]{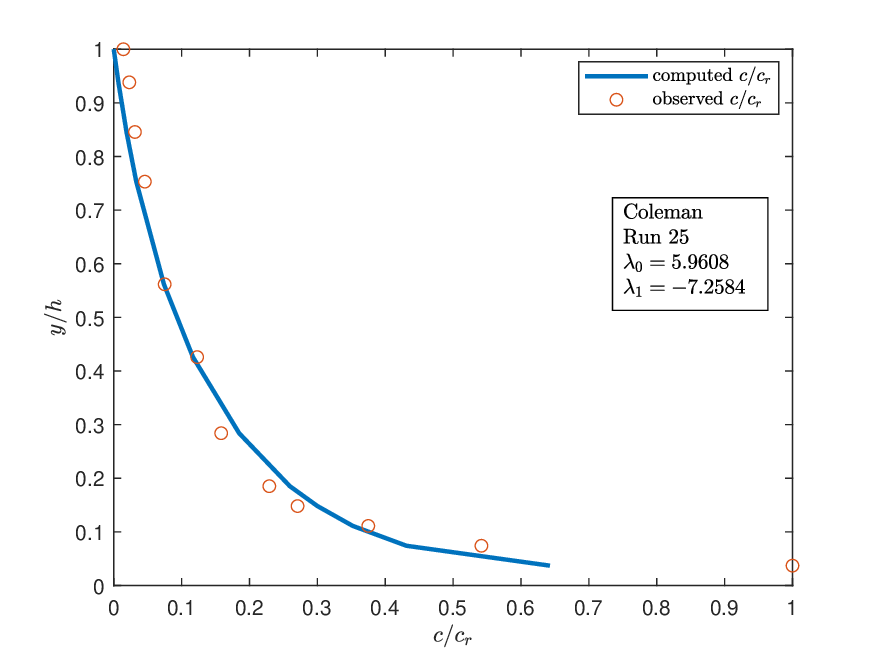} ~~~
		\includegraphics[width=0.45\textwidth]{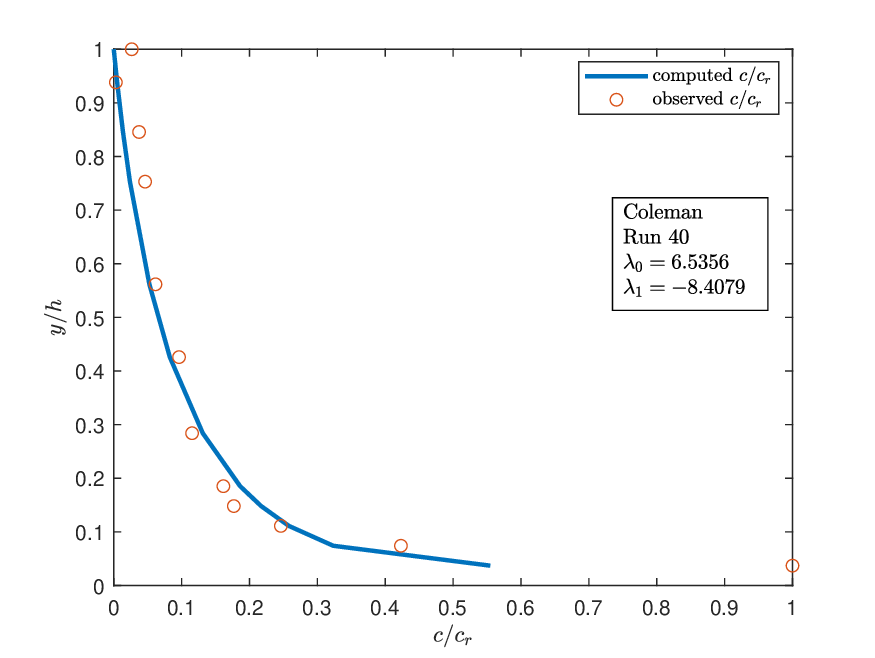} ~~~\\	
		\includegraphics[width=0.45\textwidth]{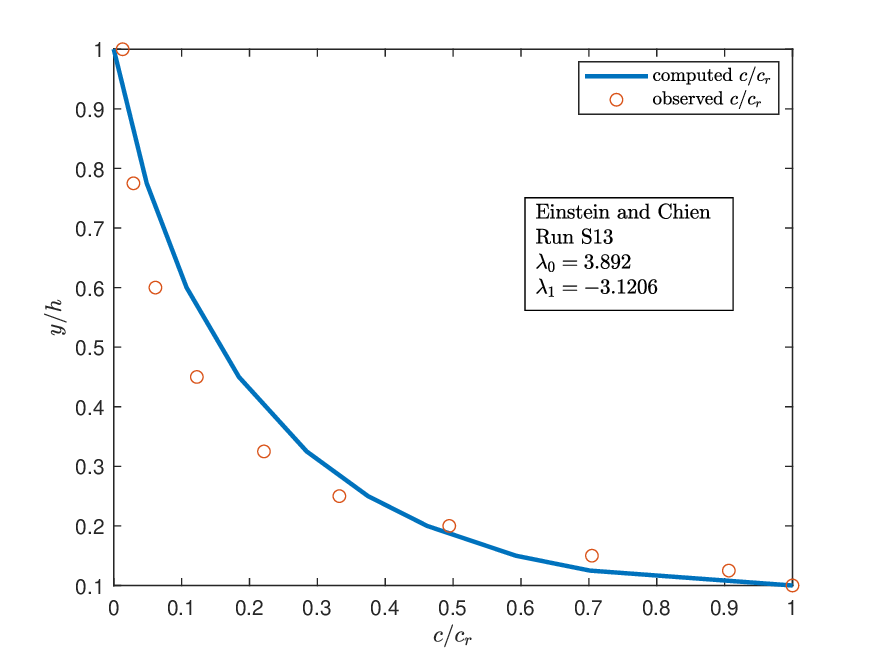} ~~~
		\includegraphics[width=0.45\textwidth]{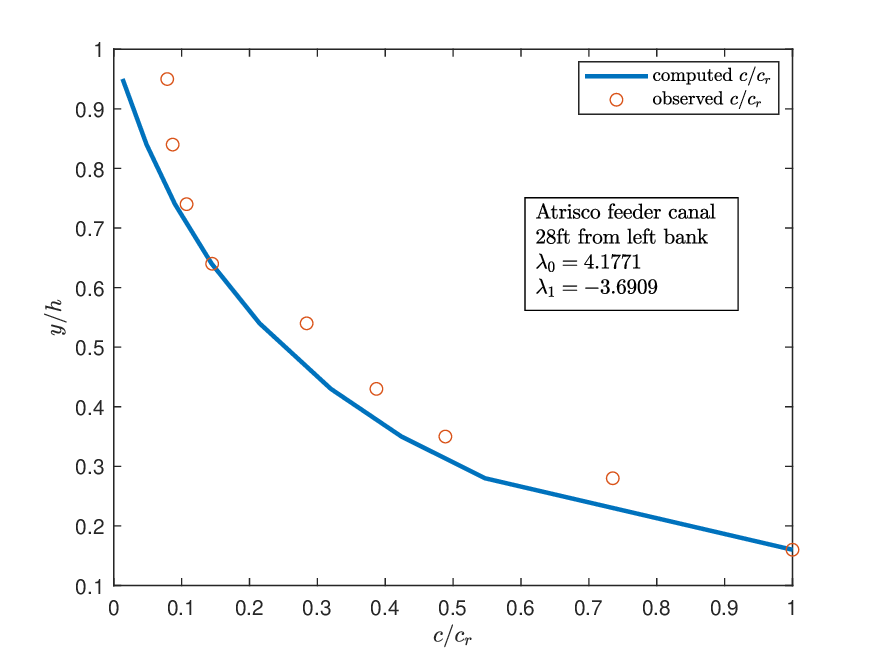} ~~~\\
		\includegraphics[width=0.45\textwidth]{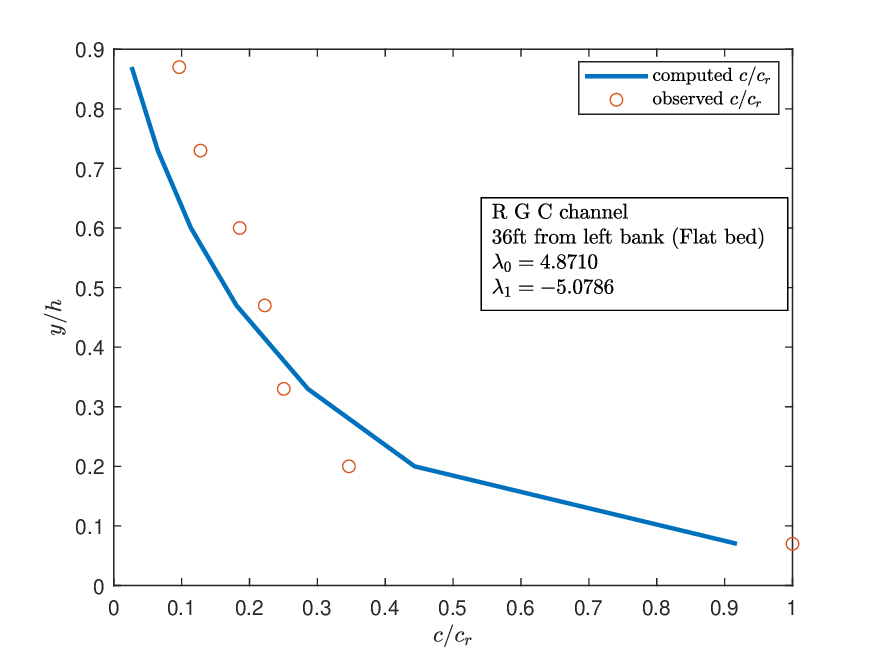} ~~~ 
		\includegraphics[width=0.45\textwidth]{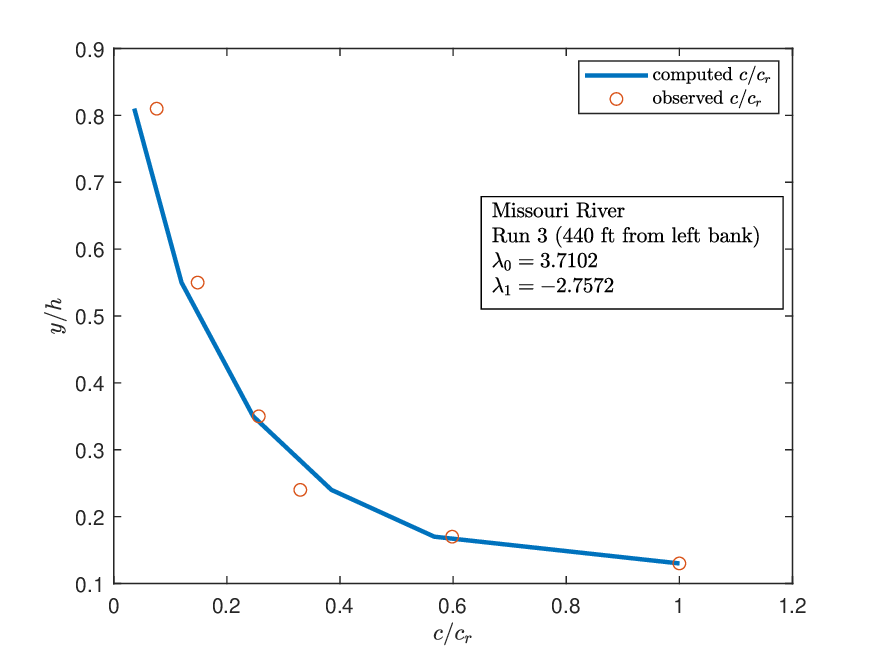} ~~~
		\caption{Dimensionless concentration ($ c/c_r $) with respect to depth ratio ($ y/h $)}
		\label{conc_ht}
	\end{figure}

\begin{figure}[]
		\includegraphics[width=0.45\textwidth]{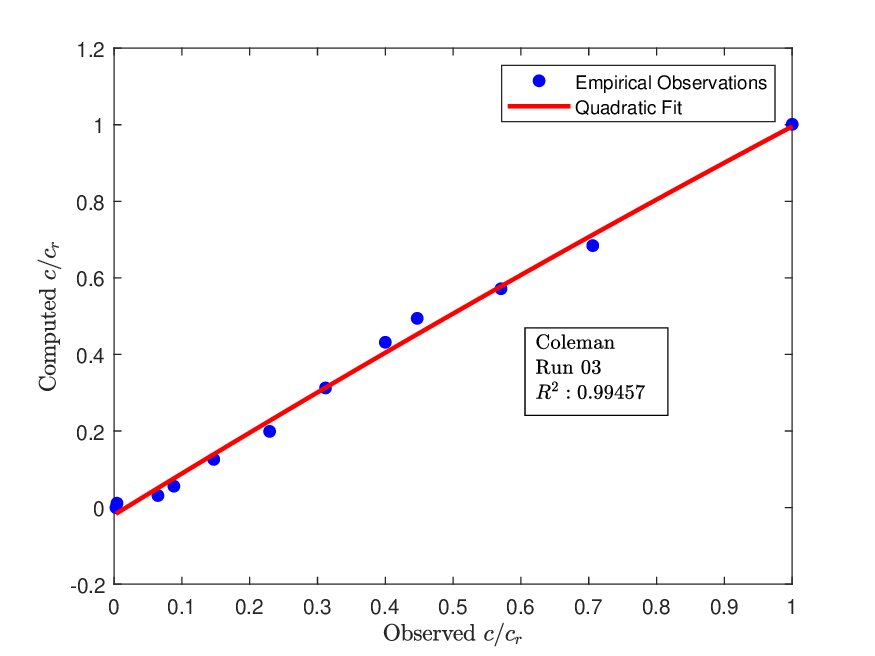} ~~~
		\includegraphics[width=0.45\textwidth]{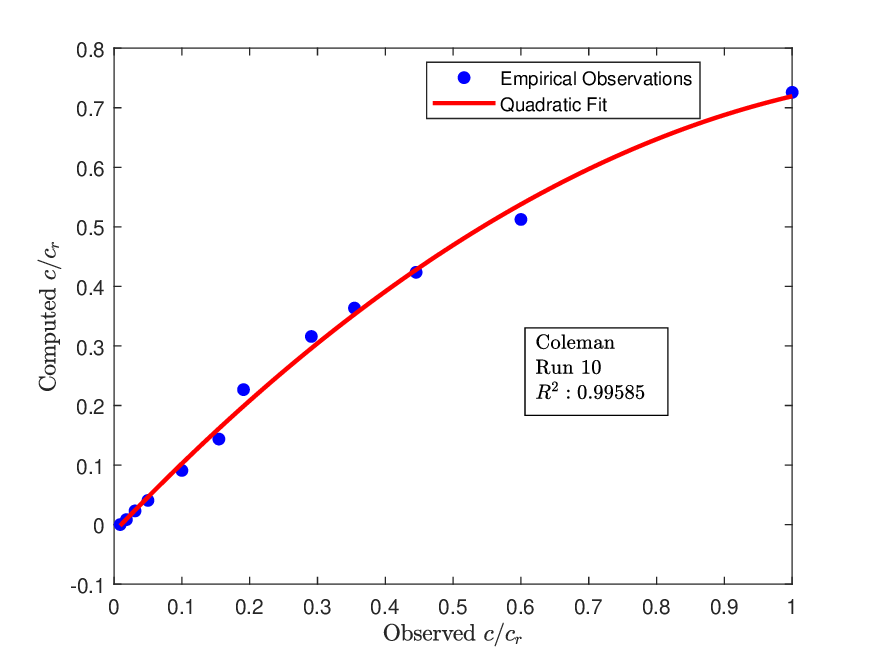} ~~~\\
		\includegraphics[width=0.45\textwidth]{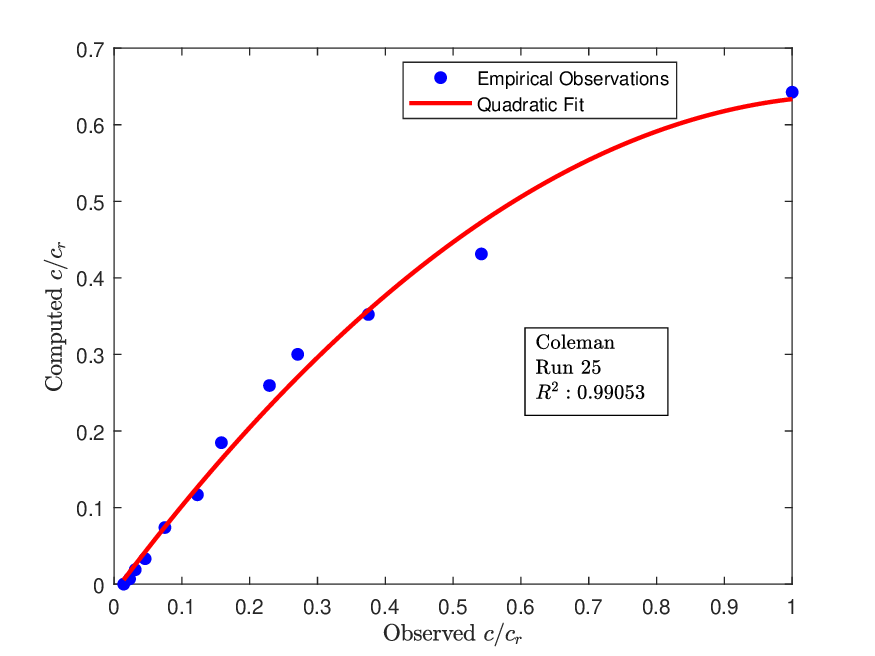} ~~~
		\includegraphics[width=0.45\textwidth]{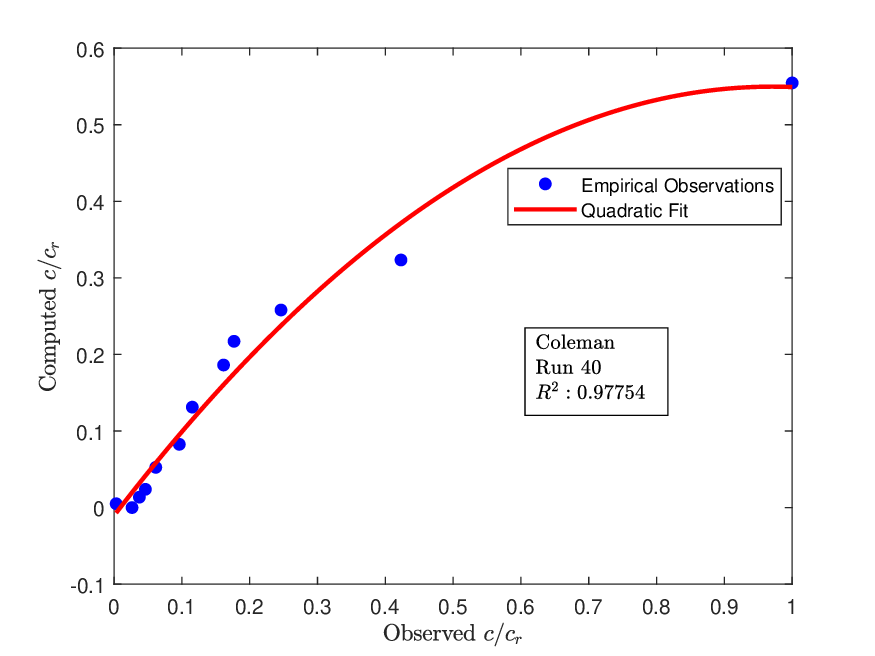} ~~~\\	
		\includegraphics[width=0.45\textwidth]{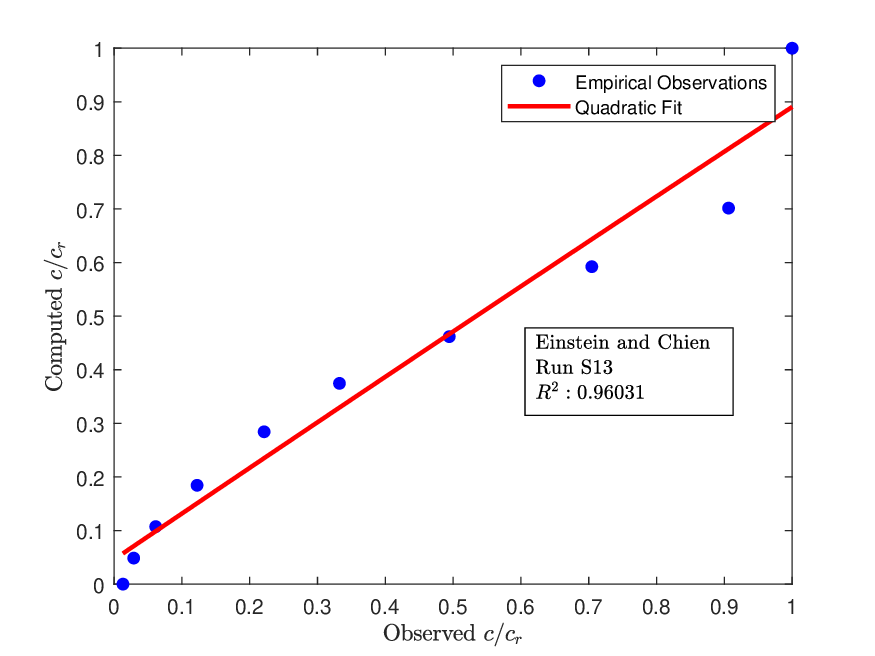} ~~~
		\includegraphics[width=0.45\textwidth]{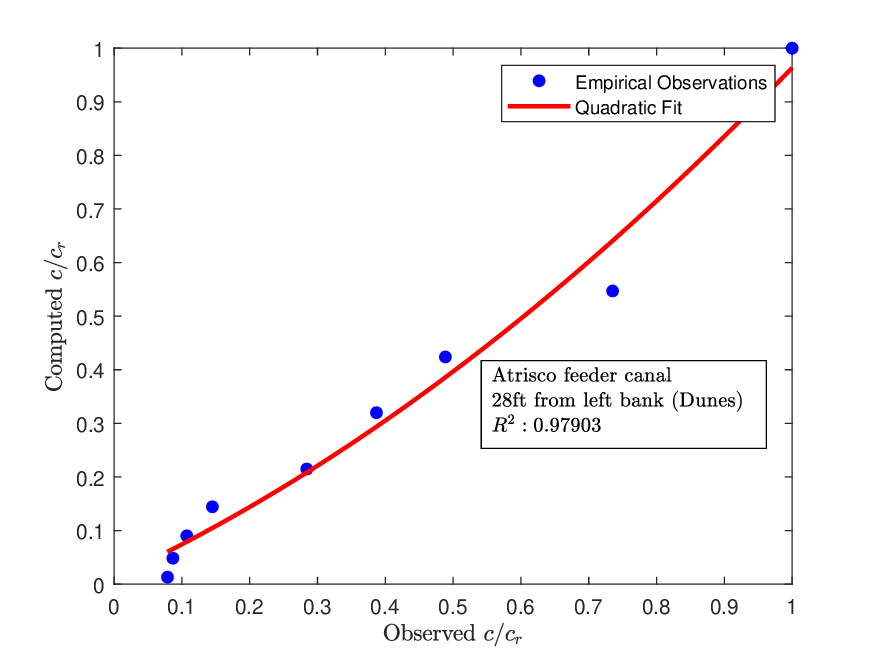} ~~~\\
		\includegraphics[width=0.45\textwidth]{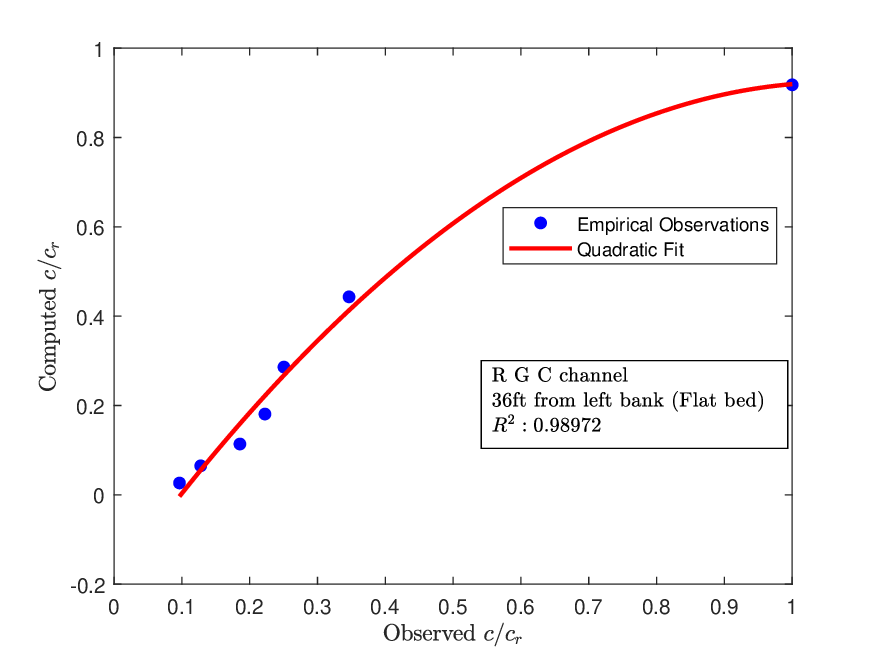} ~~~
		\includegraphics[width=0.45\textwidth]{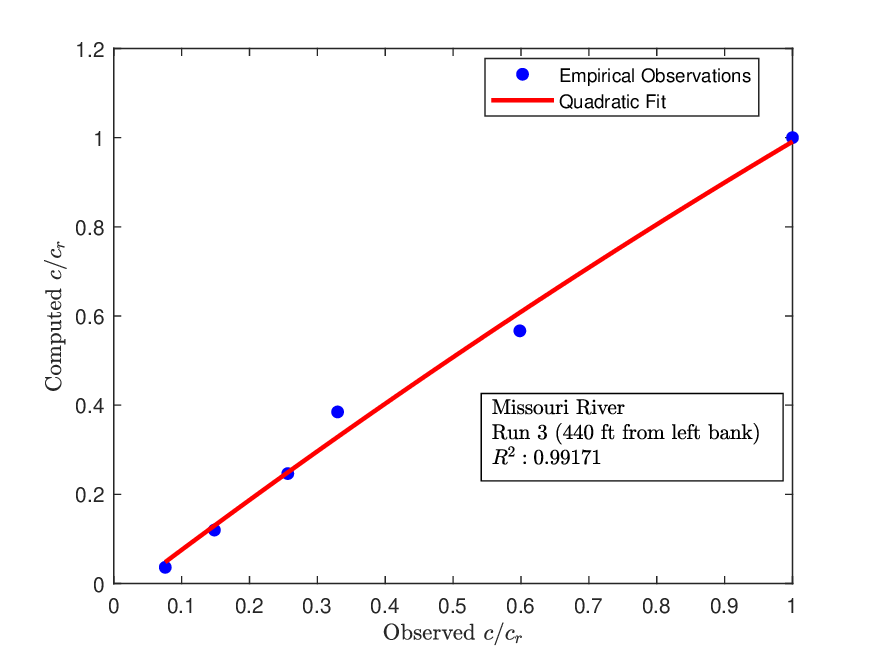} ~~~
		\caption{Regression analysis for computed $ c/c_r $ with respect to observed $ c/c_r $}
		\label{quad_fit}
	\end{figure}

	In order to justify the authenticity and accuracy of the FDE-based SSC model represented by eq. (\ref{2.24}), we have used the experimental data sets from \cite{co} and \cite{ec} along with field data from \cite{mc}. For each dataset, the reference level $ y_r $ is set as the lowest recorded height with  available concentration measurement to prevent overestimation or underestimation of the reference height derived from deterministic methods \citep{mg,mgk}. The experiment performed by \cite{ec} was done in a laboratory prepared with painted steel sheet with a width of 1.006 ft, length of 12.19 ft and a depth of 1.17 ft. In his experiment, 29 runs were performed with fine, medium sized and coarse type of sand particles placed at the bottom of the channel. The accuracy of the current model is assessed by using Run S13 data. \\
	\hspace*{0.2in} The experimental data of \cite{co} was recorded by using a narrow rectangular open channel having  length 1.5 cm, width 35.6 cm and depth 17.2 cm. A total of 40 runs were carried out with particle sizes varying from fine (0.105 mm diameter) to medium (0.2 mm diameter) and coarse (0.42 mm diameter). Runs 3, 10, 25 and 40 are chosen for the purpose of validation of the proposed concentration model. One can also find the details of the data in \cite{l,aks}.  \\
	\hspace*{0.2in} Other than the experimental data, field data were chosen from Rio Grande conveyance channel (RGC) (36 feet away from the left bank with flat channel bed and particle size of 0.202 mm), Atrisco feeder canal (AFC) (28 feet away from left bank with particle size of 0.235 mm and dunes as bed form) and Run 3 from Missouri river (440 feet away from left bank with dunes as bed form). One can find the details of the data in \cite{mc}. \\
	\hspace*{0.2in} Firstly, the computed hypothetical and the observed general cdfs identified by eqns. (\ref{2.15}) and (\ref{2.23}), respectively, are validated with the experimental and field data sets considered for this study. The parameter $ a $ is obtained by fitting the hypothetical cdf with the generalized cdf for a given set of data as shown in Fig. \ref{cdf_fit}. 
	Next, for finding the appropriate expression for concentration distribution determined through eq. (\ref{2.24}), we have plotted the dimensionless concentration $ c/c_r $ against $ y/h $ for the chosen data sets described above. From Fig. \ref{conc_ht}, one can find that our proposed model agrees well enough with both the experimental and field data. Moreover, from Fig. \ref{conc_ht}, we can tell that the expression which will rightly determine the concentration distribution of suspended sediment particles will be given by
	\begin{equation}
		\hat{c} = \frac{1}{\lambda_1} \Bigg[-\lambda_0 + \sqrt{\lambda_0^2 + \lambda_1 \Big(1 - \Big(\frac{y-y_r}{h-y_r}\Big)^a\Big) \exp\Big(- \frac{y-y_r}{h-y_r}\Big)^a\Big)2\sqrt{2}e^{1/2}} \Bigg].  \label{3.32}
	\end{equation} 	
	To check how well our estimated values obtained from (\ref{3.32}) matches with the observations from the selected data sets, we have done regression analysis of the observed and the computed values of the dimensionless concentration $ \hat{c} = c/c_r $. Fig. \ref{quad_fit} illustrates the validity of our proposed model represented by (\ref{2.24}) as we get comparatively high values of regression coefficient for the experimental data sets of \cite{co}, followed by the data from \cite{ec} and the field data from \cite{mc}. The regression coefficients ranges from 0.9958 to 0.9603, which depends on the type of data and the sediment particle size. \\ 
	\hspace*{0.2in} For further validation, we have compared the present concentration model (\ref{2.24}) with the existing entropy-based models with zero surface concentration given by (\ref{1.3}), (\ref{1.4}), (\ref{1.5}) and the deterministic Rousian model defined by (\ref{1.1}). For this purpose, we have computed different errors for both the dimensionless and dimensional concentration to show that our model has acceptable range of error as compared to other models. The types of errors  considered in this study for comparison purpose have been defined in \cite{aks}. The errors can be described as follows:
	\begin{itemize}
		\item Relative error ($ R_e $) = $\frac{1}{n}\sum_{k=1}^{n}\frac{|\hat{c}_{k(comp)}-\hat{c}_{k(obs)}|}{\hat{c}_{k(obs)}} $
		\item Sum of relative squared error (E1) = $\sum_{k=1}^{n} \frac{(\hat{c}_{k(comp)}-\hat{c}_{k(obs)})^2}{\hat{c}_{k(obs)}^2} $ 
		\item Root-mean-square error (RMSE) = $\mathlarger{\sqrt{\frac{1}{n}\sum_{k=1}^{n}(\hat{c}_{k(comp)}-\hat{c}_{k(obs)})^2}}$
		\item Mean absolute standard error (MASE) = $ \frac{1}{n}\sum_{k=1}^{n} M_i $\\
		where \[M_i = \begin{cases}
			\frac{\hat{c}_{k(comp)}}{\hat{c}_{k(obs)}} & \text{if} ~ \hat{c}_{k(comp)} > \hat{c}_{k(obs)} \\ 
			\frac{\hat{c}_{k(obs)}}{\hat{c}_{k(comp)}} & \text{if} ~ \hat{c}_{k(comp)} < \hat{c}_{k(obs)}
		\end{cases}\]
		\item Sum of logarithmic deviation error (E2) = $ \sum_{k=1}^{n} (\log |\hat{c}_{k(comp)}| - \log |\hat{c}_{k(obs)}|)^2 $
	\end{itemize} 
	where $ n $ refers to the total count of data points in a particular set of data, $ \hat{c}_{k(comp)} $ and $ \hat{c}_{k(obs)} $ denote the computed and observed concentration of sediment at the $ k^{th} $ data point. Since we have considered zero surface concentration, the last two errors have been computed by ignoring the data point at the surface to get finite values. 
	The results are shown in Tables \ref{tab1} and \ref{tab2} for dimensionless $ \hat{c} $ and dimensional concentration $ c $, respectively. The errors are found to be within acceptable limits. \\
	\hspace*{0.2in} Additionally, specific datasets are selected to evaluate and compare the error magnitudes of the developed model against those of the extant SSC models, as presented in Tables \ref{tab3} and \ref{tab4}. From Table \ref{tab3}, for Missouri River Run 03 of \cite{mc} data, Run 25 of \cite{co} data and Run S13 of \cite{ec} data, the proposed model is found to have the lowest $ R_e $, E1, RMSE and MASE for most of the other models where zero surface concentration was assumed, proving its superiority over those models. Furthermore, in order to shed light on the advantages of our proposed model over the model based on R$ \acute{e} $nyi entropy proposed by \cite{kms} and the fractional Wang entropy-based model introduced by \cite{aks}, the mean error $ \mu(\epsilon) $ and standard deviation of error $ \sigma(\epsilon) $ are compared for Runs 10, 25 and 40 of \cite{co} data and Missouri River Run 03 of \cite{mc} data. 
	The current FDE-based SSC model produces smaller mean error values $ \mu(\epsilon) $ compared to the R$\acute{e}$nyi and fractional Wang (FW) entropy-based models, which were designed to represent more realistic scenarios with non-zero surface concentration (see Table \ref{tab4}). Moreover, Table \ref{tab4} also shows that the standard deviation of error $ \sigma(\epsilon) $ is influenced by the type of dataset used rather than the model itself. Therefore, both the mean and standard deviation of errors needs to be considered for comparison. This reinforces the point that assuming non-zero surface concentration does not necessarily lead to improved results, as evidenced by the values of $ \mu(\epsilon) $ and $ \sigma(\epsilon) $ displayed in Table \ref{tab4}. It also highlights the superiority and strong predictive accuracy of the model (\ref{2.24}) proposed in this study.  
	\begin{landscape}
		\begin{table}[h]
			\small
			\centering
			\caption{Mean error $ \mu(\epsilon) $, Standard deviation of error $ \sigma(\epsilon), $ Relative error ($ R_e $), Sum of squared relative error (E1), Root-mean-square error (RMSE), Mean-absolute-standard error (MASE) and sum of logarithmic deviation error (E2) for validation of estimated dimensionless concentration $ \hat{c} $ with chosen data sets }
			\vspace*{0.4in}
			\begin{tabular}{ l c c c c c c c c}
				\hline
				\textbf{Data set} & ~ \textbf{Sample size} &~  $ \mathbf{\mu(\epsilon)} $ \textbf{(g/l)} &~ $ \mathbf{\sigma(\epsilon)} $\textbf{(g/l)}  &~ \textbf{E1} &~ \textbf{RMSE} &~ {$ \mathbf{R_e} $} &~ \textbf{MASE} &~ \textbf{E2}\\
				\hline \hline
				\textit{Coleman data} &~ &~ &~ &~ &~ &~ &~ &~ \\
				Run 03 &~ 12 &~ 0.0192 &~ 0.0161 &~ 4.134 &~ 0.0698 &~ 0.3346  &~ 1.3471 &~ 1.7321 \\
				Run 10 &~ 12 &~ 0.0425 &~ 0.0764 &~ 1.5529 &~ 0.0991 &~ 0.24337 &~ 1.2588 &~ 0.9289\\
				Run 25 &~ 12 &~ 0.0533 &~ 0.0999 &~ 1.9572 &~ 0.1283 &~ 0.28894 &~ 1.4301 &~ 2.1183\\
				Run 40 &~ 12 &~ 0.0612 &~ 0.1236 &~ 2.4021 &~0.1566 &~ 0.35531 &~ 1.4824 &~ 2.2334\\
				\textit{Einstein \& Chien data} &~ &~ &~ &~ &~ &~ &~ &~ \\
				Run S13 &~ 10 &~ 0.0595 &~ 0.0599 &~ 2.4497 &~ 0.1347 &~ 0.3792 &~ 1.3214 &~ 2.2396\\
				\textit{McQuivey data} &~ &~ &~ &~ &~ &~ &~ &~ \\
				AFC (Dunes) &~ 9 &~ 0.0568 &~ 0.0570 &~ 1.0908 &~ 0.1333 &~ 0.2498 &~ 1.7846 &~ 3.8303\\
				RGC (Flat Bed) &~ 7 &~ 0.0658 &~ 0.0215 &~ 1.0596 &~ 0.1494 &~ 0.3280 &~ 1.7138 &~ 2.5073\\
				Missouri (Run 3) &~ 6 &~ 0.0274 &~ 0.0199 &~ 0.34097 &~ 0.0789 &~ 0.1619 &~ 1.2651 &~ 0.6172\\
				\hline
			\end{tabular}
			\label{tab1}
		\end{table}	
	\end{landscape}
	
	\begin{landscape}
		\begin{table}[h]
			\small
			\centering
			\caption{Relative error ($ R_e $), Sum of squared relative error (E1), Mean-absolute-standard error (MASE), Sum of logarithmic deviation error (E2) and Root-mean-square error (RMSE) for validation of estimated concentration $ c $ with chosen data sets}
			\vspace*{0.4in}
			\begin{tabular}{ l c c c c c  }
				\hline
				\textbf{Data set} &~ \textbf{Sample size} &~ {$ \mathbf{R_e} $} &~ \textbf{E1} &~  \textbf{MASE} &~\textbf{E2} \\
				\hline \hline
				\textit{Coleman data} &~ &~ &~ &~ &~ \\
				Run 03 &~ 12 &~ 0.2596 &~ 1.9979 &~ 1.4518 &~ 2.5665 \\
				Run 10 &~ 12 &~ 0.2433 &~ 1.5524 &~ 1.2586 &~ 0.9275 \\
				Run 25 &~ 12 &~ 0.2891 &~ 1.9585 &~ 1.4306 &~ 2.1221 \\
				Run 40 &~ 12 &~ 0.3563 &~ 2.4165 &~ 1.4833 &~ 2.2396 \\
				\textit{Einstein \& Chien data} &~ &~ &~ &~ &~ \\
				Run S13 &~ 10 &~ 0.3794 &~ 2.4525 &~ 1.3217 &~ 0.9244 \\
				\textit{McQuivey data} &~ &~ &~ &~ &~ \\
				\cite{mc} AFC (Dunes) &~ 9 &~ 0.2497 &~ 1.0905 &~ 1.7842 &~ 3.8283 \\
				\cite{mc} RGC (Flat Bed) &~ 7 &~ 0.3280 &~ 1.0596 &~ 1.7136 &~ 2.5062 \\
				\cite{mc} Missouri (Run 3) &~ 6 &~ 0.1620 &~ 0.3411 &~ 1.2652 &~ 0.6176 \\
				\hline
				
			\end{tabular}
			\label{tab2}
		\end{table}	
	\end{landscape}
	
	\begin{landscape}
		\begin{table}[h]
			\small
			\centering
			\begin{threeparttable}
				\caption{Comparison of the present model with other vertical profile of SSC model}
				\vspace*{0.4in}
				\begin{tabular}{ l c c c c}
					\hline
					\textbf{SSC models} &~ \textbf{Current model}  &~ \textbf{Tsallis model} &~ \textbf{Shannon (Choo) model} &~  \textbf{Rousian model} \\
					\hline \hline
					Data source &~ \textit{Missouri River Run 03 of McQuivey data} &~ &~ &~  \\ 
					$ R_e $ &~ 0.1619* &~ 0.4286 &~ 0.4136 &~ 0.3103 \\
					E1 &~ 0.3410* &~ 1.4206  &~ 1.4193 &~ 1.0278\\
					RMSE &~ 0.0789 &~ 0.1849 &~ 0.1529 &~ 0.0607* \\
					MASE &~ 1.2651* &~ 1.4401 &~ 1.4600 &~ 2.1839 \\
					Data source &~ \textit{Run 25 of Coleman data} &~ &~ &~ \\ 
					$ R_e $ &~ 0.2890* &~ 0.5499 &~ 0.6317 &~ 0.8642 \\
					E1 &~ 1.9572* &~ 4.6198 &~ 5.9007 &~ 9.8640 \\
					RMSE &~ 0.1283* &~ 0.1439 &~ 0.1307 &~ 0.1823 \\
					MASE &~ 1.4301* &~ 1.5508 &~ 1.6116 &~ 5.5690 \\
					Data source &~ \textit{Run S13 of Einstein \& Chien data} &~ &~ &~ \\ 
					$ R_e $ &~ 0.3792* &~ 0.5511 &~ 1.2769 &~ 0.5555 \\
					E1 &~ 2.4497* &~ 6.1399 &~ 31.3755 &~ 3.8921 \\
					RMSE &~ 0.1347 &~ 0.1439 &~ 0.1307* &~ 0.1823 \\
					MASE &~ 1.3214* &~ 1.9023 &~ 2.3154 &~ 3.1771 \\
					\hline			 
				\end{tabular}
				\label{tab3}
				\begin{tablenotes}
					\footnotesize
					\item Note: * indicates the minimum values of different types of error for each data source.
				\end{tablenotes}
			\end{threeparttable}
		\end{table}	
	\end{landscape}
	
	\begin{table}[h]
		\small
		\centering
		\begin{threeparttable}
			\caption{Mean error $ \mu(\epsilon) $ and standard deviation of error $ \sigma(\epsilon) $ to compare proposed FDE model with R$\acute{e}$nyi model \citep{kms} and Fractional Wang (FW) \cite{aks} model}
			\vspace*{0.4in}
			\begin{tabular}{ l c c c }
				\hline
				\textbf{SSC models} &~ \textbf{FDE model} &~ \textbf{R$\mathbf{\acute{e}}$nyi model} &~ \textbf{FW model}  \\
				\hline \hline
				Data source (Field)  &~ &~ &~  \\
				McQuivey data: &~ &~ &~  \\
				\textit{Missouri River Run 03} &~ &~ &~  \\
				$ \mu(\epsilon) $ &~ 0.0274* &~ 0.1486 &~ 0.1215  \\ 
				$ \sigma(\epsilon) $ &~ 0.0199* &~ 0.1617 &~ 0.1397 \\
				Data source (Experimental)  &~ &~ &~  \\
				Coleman data:  &~ &~ &~  \\
				\textit{Run 10} &~ &~ &~  \\
				$ \mu(\epsilon) $ &~ 0.0425* &~ 0.0632 &~ 0.0568 \\ 
				$ \sigma(\epsilon) $ &~ 0.0764 &~ 0.0755 &~ 0.0564* \\
				\textit{Run 25} &~ &~ &~ \\
				$ \mu(\epsilon) $ &~ 0.0533* &~ 0.1366 &~ 0.0728 \\ 
				$ \sigma(\epsilon) $ &~ 0.0999 &~ 0.0700* &~ 0.1100 \\
				\textit{Run 40} &~ &~ &~ \\
				$ \mu(\epsilon) $ &~ 0.0612* &~ 0.1281 &~ 0.0957 \\ 
				$ \sigma(\epsilon) $ &~ 0.1236 &~ 0.0846* &~ 0.0999 \\
				\hline		
			\end{tabular}
			\label{tab4}
			\begin{tablenotes}
				\footnotesize
				\item Note: * represents the lowest mean and standard deviation of error across various data sources.
			\end{tablenotes}
		\end{threeparttable}
	\end{table}	
	\section{Conclusion}
	The fractional differential entropy (FDE) remains mostly unexplored for describing the uncertainty in the distribution of physical phenomena. Therefore, this poses an open challenge to test the suitability of FDE for different types of physical applications. This study presents a simplistic model in terms of FDE for describing the vertical profile of suspended sediment concentration (SSC) distribution along the vertical column of an open channel turbulent flow from a chosen reference level $ y = y_r $ to the flow surface $ y = h $. For simplicity of the model, we have assumed zero concentration at the water surface. However, for better description of realistic situations, we have taken the concentration from $ y_r $ instead of the channel bed ($ y=0 $). This reference level is taken as the lowest height with available concentration measurement for a given data. \\
	\hspace*{0.2in}In this study, we have demonstrated the superiority and enhanced prediction accuracy of the current model over existing deterministic Rousian model and entropy-based SSC models \citep{cs, ch, r} that assume zero surface concentration. This was achieved through error analysis and regression analysis of the computed and observed values. The errors in case of our proposed model were found to be minimum for most of the chosen data sets, confirming the reliability of the FDE-based SSC model. \\ 
	\hspace*{0.2in} The study also shows that the mean error $ \mu(\epsilon) $ of the current FDE-based model is lower than that of the R$\acute{e}$nyi model proposed by \cite{kms} and the fractional Wang entropy (FW) model defined by \cite{aks}. This underscores the significance of the current model among existing models with non-zero concentration at the surface. This indicates that the assumption of zero surface concentration is not necessarily a limitation in enhancing model accuracy. A Type I distribution of SSC is assumed in this study to reduce the complexity of the model. However, the formulation of Type II and Type III SSC distributions within the framework of FDE remains an unresolved issue, offering a promising avenue for future research. 
	\section*{Data availability statement}
	The sources of all the data used in this study are cited in the bibliography section.
	\section*{Disclosure statement}
	On behalf of all authors, the corresponding author states that there is no conflict of interest.

\end{document}